\documentclass[preprint,12pt]{elsarticle}

\usepackage{amsmath, amssymb, fullpage, graphics}
\usepackage{caption, graphicx, fancyhdr, fancybox}
\usepackage{color}

\renewcommand\baselinestretch{1.12}

\journal{Elsevier}

\newcommand{\bU}{\mathbf{U}}
\newcommand{\bD}{\mathbf{D}}
\newcommand{\bv}{\mathbf{v}}

\begin{document}
	
\begin{frontmatter}
		
\title{Formation and evolution of roll waves in a shallow free surface flow of a power-law fluid down an inclined plane}

\author[label1,label2]{Alexander Chesnokov}
\ead{chesnokov@hydro.nsc.ru}
		
\address[label1]{Lavrentyev Institute of Hydrodynamics SB RAS, 15 Lavrentyev Ave., Novosibirsk 630090, Russia}
\address[label2]{Novosibirsk State University, 1 Pirogova Str., Novosibirsk 630090, Russia}

\begin{abstract}
A laminar flow of a thin layer of mud down an inclined plane under the action of gravity is considered. The instability of a film flow and the formation of finite amplitude waves are studied in the framework of both two-dimensional governing equations of a power-law fluid and its depth-averaged hyperbolic simplification. Conditions of roll waves existence for these models are formulated in terms of Whitham criterion. Numerical calculations of the free surface evolution and the roll waves development are performed. It is shown that the roll waves amplitude obtained by the 2D equations is slightly larger than for the 1D model. Moreover, for certain flow parameters, small perturbations of the basic solution grow for the 2D equations and decay for the depth-averaged model. A two-parameter class of exact piecewise-smooth solutions of the 1D model is obtained and a comparison with a numerical solution is made. In the region of these parameters, diagrams of the roll waves existence are constructed.
\end{abstract}


\end{frontmatter}

\section{Introduction} 

The gravity-driven flow of a power-law liquid film down an inclined plane is an important problem with many industrial applications (chemical engineering, coating processes) and natural hazards (mud flows, debris flow impact). Steady flows of thin layers of Newtonian or non-Newtonian fluid can become unstable for sufficiently high slopes against long-wavelength infinitesimal perturbations. Such long-wave instability frequently develops into a series of progressing bores, known as roll waves. The control of this instability concerns many technologies. For instance, a uniform flow thickness is required in coating processes and, consequently, the instability should be avoided. On the other hand, in chemical engineering such waves can be useful since they enhance heat and mass exchange. In environmental hydraulics, muddy intermittent flows are frequently encountered in mountainous regions, especially after torrential rains. Moreover, these type of waves can also be observed as an event following volcano eruptions. 

Since the remarkable work of Kapitza \cite{Kapitza_1948}, wave evolution of liquid films flowing down an inclined plane has attracted a considerable number of studies \cite{Alekseenko_1994, Chang_2002}. Direct numerical simulation of such flows is difficult and requires large computational costs. Therefore, depth-averaged models are widely used to describe and study the film flow evolution. Existing theoretical models of thin layer motion of Newtonian and non-Newtonian fluid are based on simplifications of the governing equations obtained from vertical averaging. The procedure is much like the von K\'{a}rm\'{a}n--Pohlhausen power integral method used in the classical theory of boundary layers, and proceeds by first prescribing a fixed vertical structure to the flow. Then the governing equations can be vertically integrated to remove entirely one of the spatial variables. In this case the shear stress is replaced by a drag term as in the Saint-Venant model used in hydraulics. For the same reasons, one uncovers roll-wave instability, the mathematical theory of which was developed by Dressler \cite{Dressler} for shallow flows in inclined channels. Shkadov \cite{Shkadov_1967} was the first who applied this approach to obtain a two-equation model for laminar film flows of a viscous liquid. Many other researchers adopted his idea to more complex geometries or non-Newtonian fluids. Ng and Mei \cite{Ng_Mei} investigated the laminar flow of a shallow layer of mud by using a shear thinning power-law rheological model and applied a theory of roll waves for such flows. The behaviour and stability of a thin layer of non-Newtonian fluids accounting for the presence of yield stress was studied by Liu and Mei \cite{Liu_Mei_1994} for the Bingham rheology. Roll waves in mud modelled as  a viscoplastic fluid with the Herschel–Bulkley constitutive law was considered by  Balmforth and Liu \cite{Balmforth_2004}. 

Numerous subsequent works are aimed at applying the theory of roll waves for various hydrodynamic problems, as well as constructing depth-averaged models, their theoretical analysis and use for modelling the evolution of finite amplitude waves. For instance, in \cite{Chesn_2018} a fluid flow in a long tube with elastic walls is considered and shown that for certain parameters of the flow small perturbations at the inlet section of the tube give rise to roll waves. The stability of two-layer flows in open and closed channels, as well as in a Hele--Shaw cell, is considered in \cite{Boutounet_2012, Boudlal_2008} and \cite{Chesn_Liap_St_2017}, where the possibility of the roll waves formation in such systems is shown. In \cite{BL02, Boudlal_2012} the modulation equations for free surface shallow flows are derived and the stability criterion of roll waves is formulated in terms of hyperbolicity of these equations. Starting with an expansion of the velocity field in terms of polynomial test functions and using a weighted-residual technique based on a classical long-wavelength expansion more complex depth-averaged models for Newtonian and power-law fluid are derived and verified in \cite{Ruyer_2000, Ruyer_2012}. Consistent shallow water equations for the flow of thin films of power-law fluid are developed in \cite{Noble_2013} through a second-order asymptotic expansion of the fluid velocity field and fluid strain. Roll waves in flows of thin films of a viscous liquid and shear shallow water flows in inclined channels are studied in works \cite{Richard_2016} and \cite{Ivanova_2017} in the framework of advanced three-equation models. It is shown that these hyperbolic governing equations more accurately reproduce the experimental data in comparison with the widely used two-equation models.

Last years, considerable attention has been paid to the study of the stability and roll waves development in shallow flows of non-Newtonian fluids. Spatial evolution of a small disturbance in an open-channel flow of a power-law fluid at non-uniform initial conditions, up to the occurrence of roll waves in mild and steep slope channels is investigated in \cite{Campomaggiore_2016}. With reference to the propagation of mud flows described by means of a power-law rheology, the potentialities of different approximations of the linearised version of the two-equation  model and the simplified kinematic-wave model are studied in \cite{Cristo_2018, Cristo_2019}. The linear stability of shear-thinning fluid down an inclined plane in the framework of the generalized Orr-Sommerfeld equation is performed in \cite{Nsom_2018}. The influence of a prescribed superficial shear stress on the generation and structure of roll waves developing from infinitesimal disturbances on the surface of a power-law fluid layer is investigated in \cite{Pascal_2007}. The effect of oblique perturbations that propagate at an arbitrary angle to the velocity of the undisturbed fluid flow moving down an incline plane under gravity is studied in \cite{Zayko_2019}. It is shown that under certain conditions, oblique perturbations can lead to the flow instability. 

In the present paper we consider the occurrence of roll waves in a free surface shallow flow of a power-law fluid using both the two-dimensional boundary layer equations and the depth-averaged hyperbolic model proposed by Ng and Mei \cite{Ng_Mei}. A generalized theory of characteristics and the notion of hyperbolicity for integro-differential equations introduced by Teshukov \cite{Teshukov_1994} (see also \cite{Chesn_2014, Chesn_etal_2017}) allow one to apply Whitham's stability criterion to both considered models. Numerical simulations are carried out in order to confirm the achievements of the theoretical analysis as well as to compare the results of calculations using fully nonlinear models of different levels of complexity. We also construct and study a two-parameter class of exact solutions for the 1D model describing the roll waves regime. It is shown that with an appropriate choice of the critical depth and amplitude of the wave, this solution is consistent with the results of numerical calculations. Diagrams of the roll waves existence that bound all possible values of the critical depth and fluid depth are obtained. 

The paper is structured as follows. In Section 2 the governing equations are reported. Then the Whitham stability criterion is formulated in Section 3 for both 1D and 2D models. For numerical treatment of the 2D equations in Section 4, a multilayer approximation is derived in the form of 1D balance laws. Numerical simulation of the development of roll waves instability using the considered models is performed in Section 5. A two-parameter class of piecewise-smooth travelling wave solutions is studied in Section 6. Finally, conclusions are drawn in Section 7.

\section{Mathematical model}

We consider a power-law fluid flowing down an inclined plane under the action of gravity. The flow is assumed to be incompressible and the fluid density $\rho$, the angle of inclination $\theta$ and the gravity acceleration $g$ are constant. An $(x,z)$ coordinate system is defined with the $x$-axis along and the $z$-axis normal to the plane bed. The velocity field and the total pressure are denoted by $\bv=(u,w)^T$ and $p$, respectively. The dimensional governing equations read
\begin{equation}\label{eq:gov-eq}
 \begin{array}{l}\displaystyle
  \frac{\partial u}{\partial t}+u\frac{\partial u}{\partial x}+
  w\frac{\partial u}{\partial z}+\frac{1}{\rho}\frac{\partial p}{\partial x}=
  \frac{1}{\rho}\Big(\frac{\partial \tau_{xx}}{\partial x}+
  \frac{\partial \tau_{xz}}{\partial z}\Big)+g\sin{\theta}, \\[3mm]\displaystyle
  \frac{\partial w}{\partial t}+u\frac{\partial w}{\partial x}+
  w\frac{\partial w}{\partial z}+\frac{1}{\rho}\frac{\partial p}{\partial z}=
  \frac{1}{\rho}\Big(\frac{\partial \tau_{zx}}{\partial x}+
  \frac{\partial \tau_{zz}}{\partial z}\Big)-g\cos{\theta}, \\[3mm]\displaystyle
  \frac{\partial u}{\partial x}+\frac{\partial w}{\partial z}=0, \quad 
  \tau_{ij}=2\mu_n(2D_{kl}D_{kl})^{(n-1)/2}D_{ij},
 \end{array}
\end{equation}
where $D_{ij}$ are the components of strain-rate tensor $\bD=(\nabla\bv+(\nabla\bv)^*)/2$, positive constants $n$ and $\mu_n$ are the power-law index and the fluid consistency respectively. The sum convention is employed in the last formula in Eq.~(\ref{eq:gov-eq}) and, consequently, the terms $\tau_{ij}$ have the form
\[ \begin{array}{l}\displaystyle 
   \tau_{xx}=2\tau_n u_x, \quad \tau_{xz}=\tau_{zx}=\tau_n(u_z+w_x), \quad 
   \tau_{zz}=2\tau_n w_z, \\[2mm]\displaystyle
   \tau_n=\mu_n\big(2u_x^2+2w_z^2+(u_z+w_x)^2\big)^{(n-1)/2}.
 \end{array} \]
If the fluid behaviour is shear-thinning then $n<1$, whereas $n>1$ corresponds to a shear-thickening fluid. Setting $n=1$, the Newtonian case is recovered, with the consistency $\mu_n$ coinciding with the dynamic viscosity. We note that power-law fluids serve as a simple but an efficient model for a broad variety of non-Newtonian fluids. The system of equations is completed by the boundary conditions at the wall $z=0$ and at the free surface $z=h$:
\begin{equation}\label{eq:BC}
 \begin{array}{l}\displaystyle
  u\big|_{z=0}=0, \quad w\big|_{z=0}=0, \quad \frac{\partial h}{\partial t}+ 
  u\frac{\partial h}{\partial x}-w\Big|_{z=h}=0, \\[3mm]\displaystyle
  (\tau_{xx}-p)\frac{\partial h}{\partial x}-\tau_{xz}\Big|_{z=h}=0, \quad \tau_{xz}\frac{\partial h}{\partial x}+\tau_{zz}-p\Big|_{z=h}=0.
 \end{array}
\end{equation}
The surface tension is neglected here. 

System (\ref{eq:gov-eq}), (\ref{eq:BC}) admits the Nusselt-type solution
\begin{equation}\label{eq:Nus-profile}   
  u(z)=f(z)U_0, \quad f(z)=\frac{1+2n}{1+n}\bigg(1-\Big(1-\frac{z}{h}\Big)^{1+1/n}\bigg)\,,
\end{equation}
where 
\begin{equation}\label{eq:Nus-vel} 
  U_0=\frac{n}{1+2n}\bigg(\frac{\rho g\sin\theta}{\mu_n}\bigg)^{1/n}\,H_0^{1+1/n}, 
\end{equation}
and $w=0$, $p=(h-z)g\cos\theta$, $h=H_0={\rm const}$. Indeed, direct calculations show that these formulae give an exact solution to equations (\ref{eq:gov-eq}) with boundary conditions (\ref{eq:BC}). Velocity profile (\ref{eq:Nus-profile}) and relation (\ref{eq:Nus-vel}) will be used further. 

Let us now write this system in a non-dimensional form and in the shallow-water scaling. We assume that the characteristic scales of the film depth, its velocity and longitudinal wavelength are $H_0$, $U_0$ and $L_0=(g\sin\theta)^{-1}U_0^2$, respectively. The shallowness of the flow is determined by the aspect ratio $\varepsilon=H_0/L_0\ll 1$. In addition, it is assumed that the characteristic velocity $U_0$ and the flow depth $H_0$ are related by formula (\ref{eq:Nus-vel}). Then we scale the variables as follows
\[ \begin{array}{l}\displaystyle
    x=L_0x', \quad (h,z)=H_0(h',z'), \quad t=\frac{L_0}{U_0}t', \quad u=U_0u', \quad 
    w=\varepsilon U_0w',\\[3mm]\displaystyle 
    p=\rho U_0^2 p', \quad (\tau_{xx},\tau_{zz})= 
    \varepsilon\mu_n\bigg(\frac{U_0}{H_0}\bigg)^n(\tau'_{xx},\tau'_{zz}), \quad
    \tau_{xz}=\mu_n\bigg(\frac{U_0}{H_0}\bigg)^n\tau'_{xz}.
   \end{array} \]
Following \cite{Ng_Mei}, we defined the non-dimensional Froude and Reynolds numbers as
\[ Fr=\frac{U_0^2}{gH_0}=\frac{\sin\theta}{\varepsilon}, \quad   
   Re=\frac{U_0^2\rho}{\mu_n}\bigg(\frac{H_0}{U_0}\bigg)^n= 
   \frac{1}{\varepsilon K}\,, \]
where $K=(n/(1+2n))^n$. 

Then, by dropping the primes, one may rewrite the momentum equations and the dynamic boundary conditions in the form
\begin{equation}\label{eq:dim-less}
 \begin{array}{l}\displaystyle
  \frac{\partial u}{\partial t}+u\frac{\partial u}{\partial x}+
  w\frac{\partial u}{\partial z}+\frac{\partial p}{\partial x}=
  K\Big(\varepsilon^2\frac{\partial\tau_{xx}}{\partial x} +\frac{\partial \tau_{xz}}{\partial z}\Big)+1, \\[3mm]\displaystyle
  \varepsilon^2\Big(\frac{\partial w}{\partial t}+u\frac{\partial w}{\partial x}+ w\frac{\partial w}{\partial z}\Big)+ \frac{\partial p}{\partial z}= \varepsilon^2 K \Big(\frac{\partial \tau_{xz}}{\partial x}+ \frac{\partial \tau_{zz}}{\partial z}\Big)- \varepsilon\cot\theta; \\[3mm]\displaystyle
  \varepsilon^2 K\big(\tau_{xx}-p\big)\frac{\partial h}{\partial x} -K \tau_{xz}\Big|_{z=h}=0, \quad \varepsilon^2 K\Big(\tau_{xz} \frac{\partial h}{\partial x}+ \tau_{zz}\Big)-p\Big|_{z=h}=0.
 \end{array} 
\end{equation}
The rest equations in formulae~(\ref{eq:gov-eq}) and (\ref{eq:BC}) are unchanged. Neglecting terms of order $\varepsilon^2$ in Eq.~(\ref{eq:dim-less}), we obtain a simplified version of the governing equations
\begin{equation}\label{eq:m-simplified}
 \begin{array}{l}\displaystyle
  \frac{\partial u}{\partial t}+u\frac{\partial u}{\partial x}+
  w\frac{\partial u}{\partial z}+ \frac{\partial p}{\partial x}=
  K\frac{\partial \tau_{xz}}{\partial z}+1, \quad 
  \frac{\partial p}{\partial z}=-\varepsilon\cot\theta, \quad 
  \frac{\partial u}{\partial x}+\frac{\partial w}{\partial z} =0; \\[3mm]\displaystyle
  u\big|_{z=0}=0, \quad w\big|_{z=0}=0, \quad 
  h_t+uh_x-w\Big|_{z=h}=0, \quad \tau_{xz}\big|_{z=h}=0, \quad p\big|_{z=h}=0. 
 \end{array}
\end{equation}
In this approximation the fluid pressure obeys the hydrostatic law $p=(h-z)\varepsilon\cot\theta$ and the dimensionless stress tensor component $\tau_{xz}$ has the form $\tau_{xz}=|u_z|^{n-1}u_z$. Below we assume that $u_z\geq 0$. Taking this into account, Eq.~(\ref{eq:m-simplified}) can be written as
\begin{equation}\label{eq:thin-layer}
 \begin{array}{l}\displaystyle
  \frac{\partial u}{\partial t}+u\frac{\partial u}{\partial x}+
  w\frac{\partial u}{\partial z}+\alpha\frac{\partial h}{\partial x}=
  K\frac{\partial}{\partial z}\bigg(\frac{\partial u}{\partial z}\bigg)^n+ 1, \\[3mm]\displaystyle
  \frac{\partial h}{\partial t}+\frac{\partial }{\partial x} \bigg(\int_0^h u\,dz\bigg)=0, \quad u\big|_{z=0}=0,\quad \frac{\partial u}{\partial z}\Big|_{z=h}=0.
 \end{array}
\end{equation}
where
\[ \alpha=\varepsilon\cot\theta, \quad w=-\int_0^z u(t,x,z')\,dz'. \]

Let us integrate the first equation of system~(\ref{eq:thin-layer}) with respect to $z$ under assumption that the velocity profile has the form $u=f(z)\bar{u}(t,x)$, where $f(z)$ is given by formula~(\ref{eq:Nus-profile}) with $h=h(t,x)$. Profiles of $f(z)$ for different $n$ are plotted in Fig.~\ref{fig:fig_1}. Using other equations of system~(\ref{eq:thin-layer}), we obtain the following depth-averaged model~\cite{Ng_Mei}:
\begin{equation}\label{eq:av-model}
 (\bar{u}h)_t+\big(\beta \bar{u}^2h+\alpha h^2/2\big)_x= h-\bigg(\frac{\bar{u}}{h}\bigg)^n, \quad h_t+(\bar{u}h)_x=0.
\end{equation}
Here
\[ \beta= \frac{1}{h} \int_0^h f^2(y)\,dy= \frac{2(1+2n)}{2+3n}>1, \]
unknowns $\bar{u}(t,x)$ and $h(t,x)>0$ are the depth-averaged velocity and the film thickness. System (\ref{eq:av-model}) includes two parameters, $\alpha\geq 0$ and $n>0$. 

Further, we consider the generation and structure of finite amplitude waves developing from infinitesimal disturbances on the free surface in the frame of both 2D model~(\ref{eq:thin-layer}) and its depth-averaged simplification~(\ref{eq:av-model}). 

\section{Whitham criterion}

\begin{figure}[t]
 \begin{center}
  \resizebox{.48\textwidth}{!}{\includegraphics{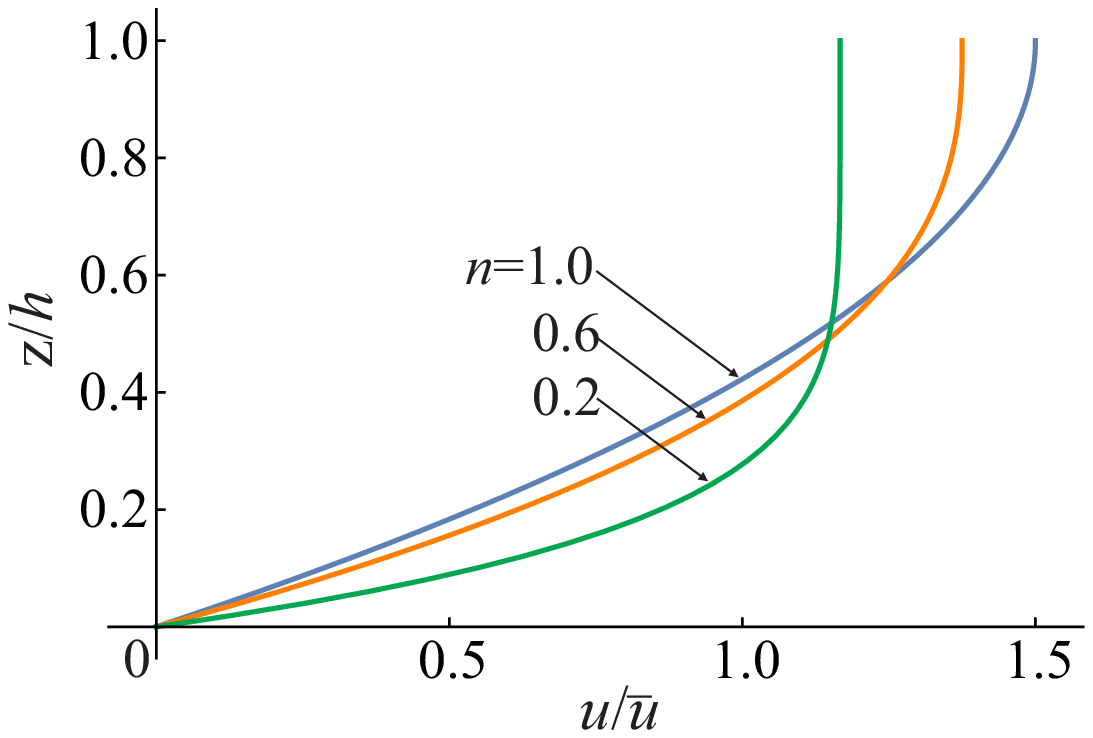}}\hfill
  \resizebox{.48\textwidth}{!}{\includegraphics{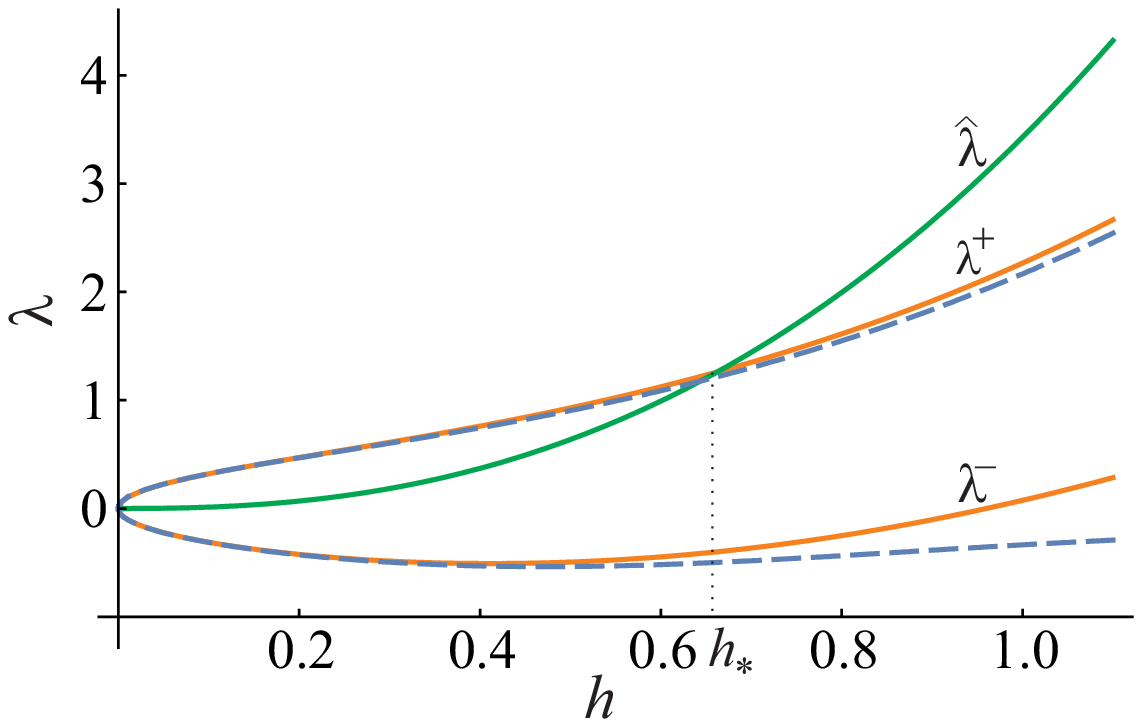}}\\[0pt]
  \parbox{.48\textwidth}{\caption{Nusselt-type velocity profiles $u/\bar{u}=f(z)$ given by formula~(\ref{eq:Nus-profile}) for power-law index $n=1$, 0.6 and 0.2.} \label{fig:fig_1}} \hfill
  \parbox{.48\textwidth}{\caption{The `frozen' $\lambda^\pm(h)$ and `equilibrium' $\hat{\lambda}(h)$ characteristic velocities for $\alpha=1$, $n=0.7$. Corresponding slopes $c^\pm(h)$ of Eq.~(\ref{eq:thin-layer}) are shown by dashed curves.} \label{fig:fig_2}} 
 \end{center}
\end{figure}

It is easy to verify that system~(\ref{eq:av-model}) is hyperbolic. There are two families of characteristics with local slopes
\begin{equation}\label{eq:char}
 \lambda^\pm=\beta\bar{u}\pm\sqrt{\beta(\beta-1)\bar{u}^2+\alpha h}.
\end{equation}
Together with Eq. (\ref{eq:av-model}) we consider more simple kinematic model where inertial effects are not taken into account. Then the first equation in (\ref{eq:av-model}) is replaced by the condition $\bar{u}=h^{1+1/n}$. Substitution of this expression in the second equation of (\ref{eq:av-model}) yields
\begin{equation}\label{eq:kinematic}
  h_t+(\varphi(h))_x=0, \quad  \varphi(h)=h^{2+1/n}.
\end{equation}
The velocity of characteristic of the kinematic-wave model (\ref{eq:kinematic}) is
\[ \hat{\lambda}=\varphi'(h)=\frac{2n+1}{n}\,h^{1+1/n}. \]
Eq.~(\ref{eq:kinematic}) is used for formulation a criterion of the roll waves existing for system~(\ref{eq:av-model}). The criterion was first described by Whitham in \cite{Whitham} for the shallow water flows in inclined open channels. This criterion is based on comparison of `frozen' characteristic velocities $\lambda^\pm(h)$ of system~(\ref{eq:av-model}) with `equilibrium' characteristic speed $\hat{\lambda}(h)$ of Eq.~(\ref{eq:kinematic}). The variables $\lambda^\pm(h)$ are obtained by substitution of $\bar{u}=\varphi(h)/h$ into the right-hand side of (\ref{eq:char}). According to Whitham criterion~\cite{Whitham} the roll waves exist in the intervals where one of inequalities is valid
\begin{equation}\label{eq:Whitham}
 \hat{\lambda}(h)>\lambda^+(h) \quad \text{or} \quad \hat{\lambda}(h)<\lambda^-(h).
\end{equation}
A typical behaviour of the `frozen' $\lambda^\pm(h)$ and `equilibrium' $\hat{\lambda}(h)$ characteristic velocities are shown in Fig.~\ref{fig:fig_2} by solid curves. The figure is obtained for $\alpha=1$ and $n=0.7$. 

Curves $\lambda=\lambda^+(h)$ and $\lambda=\hat{\lambda}(h)$ intersect at the point
\begin{equation}\label{eq:h_ast}  
  h=h_*=\bigg(\frac{\alpha n^2}{1+2n}\bigg)^{n/(2+n)}\,. 
\end{equation}
Therefore, in accordance with condition (\ref{eq:Whitham}) and formula (\ref{eq:h_ast}), the constant solution $h=h_0$, $\bar{u}=h_0^{1+1/n}$ is stable if $h_0<h_*$ and unstable otherwise. Without loss of generality, one can choose $h_0=1$. Then the stability condition takes the form 
\begin{equation}\label{eq:alpha-stab} 
  \alpha>\alpha_n=\frac{1+2n}{n^2}. 
\end{equation}
Inequality (\ref{eq:alpha-stab}) coincides with the condition obtained in \cite{Ng_Mei} based on the linear stability analysis of the constant solution $h=1$, $\bar{u}=1$. As far as we know, linear analysis and Whitham's criterion always give the same result for hyperbolic two-equation systems (see \cite{Chesn_2018, Chesn_Liap_St_2017} for instance).

A generalized theory of characteristics for integro-differential equations was introduced by Teshukov \cite{Teshukov_1994} (see also \cite{Teshukov_2004, Chesn_2014, Chesn_etal_2017}), who developed new mathematical tools for the qualitative study of such systems. As it follows from the cited works, characteristic velocities $c=c(t,x)$ of the integro-differential system (\ref{eq:thin-layer}) on a solution $u(t,x,z)$, $h(t,x)$ are determined by the equation 
\begin{equation}\label{eq:char-eq} 
  \chi(c)=1-\alpha\int_0^h \frac{dz}{(u-c)^2}=0. 
\end{equation}
It is known that for flows with a monotonic velocity profile (e.g. $0\leq u_z<\infty$) this equation has exactly two real roots $c=c^-<u(t,x,0)$ and $c=c^+>u(t,x,h)$. Let us calculate the characteristic velocities $c^\pm$ on the  Nusselt-type solution $u=f(z)h^{1+1/n}$, $h={\rm const}$ of Eq.~(\ref{eq:thin-layer}). Substitution of this solution into Eq.~(\ref{eq:char-eq}) gives an implicit equation of the form 
\begin{equation}\label{eq:char-eq-sp}  
  1- \frac{\alpha h}{(1+n)(c-u_s)c}\big(n+\, _2F_1(1,1;\kappa;u_s/c)\big)=0,   
   \quad \bigg(u_s=\kappa h^{1+1/n}, \ \kappa=\frac{1+2n}{1+n}\bigg) 
\end{equation}
which determines the characteristic velocities $c=c^\pm(h)$. Here $_2F_1(a,b;c;z)$ is the Gaussian hypergeometric function. For $\alpha=1$ and $n=0.7$ the functions $c=c^\pm(h)$ are shown in Fig.~\ref{fig:fig_2} by dashed lines. As we can see in Fig.~\ref{fig:fig_2}, the curves $\lambda=\lambda^+(h)$ and $\lambda=c^+(h)$ are located quite close. The same is true for other values of the parameters $\alpha$ and $n$. We also note that the `equilibrium' simplification of system~(\ref{eq:thin-layer}) coincides with the kinematic-wave equation~(\ref{eq:kinematic}). This allows one to apply Whitham's criterion to study the stability of the Nusselt-type solution of integro-differential system (\ref{eq:thin-layer}). To do this, we can use formulae~(\ref{eq:Whitham}), where $\lambda^\pm(h)$ should be replaced by $c^\pm(h)$. Since the function $f(z)$ is convex (its second derivative does not change sign on the interval $z\in (0,h)$ for $n\neq 1$), according to \cite{Chesn_etal_2017} the characteristic equation~(\ref{eq:char-eq}) has no complex roots (such that ${\rm Im}\,(c)\neq 0$).

Let us note, that substitution of $h=1$ and $c=\hat{\lambda}(1)=2+1/n$ in Eq.~(\ref{eq:char-eq-sp}) yields an analogue of the stability condition~(\ref{eq:alpha-stab}) in the form 
\begin{equation}\label{eq:alpha-stab-gen} 
  \alpha>\alpha_n^*=\frac{(1+2n)\alpha_n}{n+\,_2F_1(1,1;\kappa;n/(1+n))}, 
\end{equation}
where $\alpha_n$ is defined in (\ref{eq:alpha-stab}). Thus, the Nusselt-type solution $u=f(z)$, $h=1$ of Eq.~(\ref{eq:thin-layer}) is stable if condition (\ref{eq:alpha-stab-gen}) is satisfied. We note that $\alpha_n^*>\alpha_n$ for all $n>0$. For instance, $\alpha_n\approx 4.90$, $\alpha_n^*\approx5.47$ at $n=0.7$, and $\alpha_n=3$, $\alpha_n^*\approx3.50$ at $n=1$. This circumstance allows us to construct examples in which small perturbations of the Nusselt-type solution of system (\ref{eq:thin-layer}) lead to the formation of finite amplitude waves, while the corresponding constant solution of the depth-averaged model (\ref{eq:av-model}) turns out to be stable. Further, we carry out numerical calculations based on both considered models and demonstrate the Whitham criterion applicability.

\section{Multilayer approximation} 

The construction of a numerical solution to the hyperbolic system of balance laws~(\ref{eq:av-model}) does not cause difficulties and can be performed according to Godunov-type schemes. This approach is not directly applicable to integro-differential model~(\ref{eq:thin-layer}). However, a multilayer approximation of shear flow equations proposed in \cite{Teshukov_2004, Chesn_2014} allows one to treat model~(\ref{eq:thin-layer}) as a system of one-dimensional balance laws.

As it was shown in \cite{Lipatov_2004, Chesn_2014}, the evolution of the smooth solution of system~(\ref{eq:thin-layer}) can involve a gradient catastrophe. The further description of the solution is possible only in the class of discontinuous functions. It leads to the necessity to formulate this model in the form of balance laws. To do this, we present Eq.~(\ref{eq:thin-layer}) in conservative form using semi-Lagrangian coordinates. In these variables Eq.~(\ref{eq:thin-layer}) are reduced to a quasilinear integro-differential system~\cite{Teshukov_2004, Chesn_2014} 
\begin{equation}\label{eq:BL-Lagr}
\begin{array}{l}\displaystyle
u_t+uu_x+\alpha\int_0^1 H_x \, d\xi =1+ \frac{K}{H} \frac{\partial}{\partial \xi} \bigg(\frac{u_\xi}{H}\bigg)^n, \quad H_t+(uH)_x=0.
\end{array}
\end{equation}
with boundary conditions $u|_{\xi=0}=0$ and $u_\xi|_{\xi=1}=0$. Here $\xi\in [0,1]$ is Lagrangian coordinate, $H=\Phi_\xi>0$ is a Jacobian of the change of variables $z=\Phi(t,x,\xi)$, function $\Phi$ is a solution of the Cauchy problem
\[ \Phi_t+u(t,x,\Phi)\Phi_x=w(t,x,\Phi), \quad \Phi|_{t=0}=\xi h(0,x). \]
System (\ref{eq:BL-Lagr}) can be rewritten in an equivalent conservative form (see \cite{Teshukov_2004, Chesn_2014} for details) 
\begin{equation}\label{eq:ConsForm}
 \begin{array}{l}\displaystyle
  H_t+(uH)_x=0, \quad (\omega H)_t+ (u \omega H)_x= K\frac{\partial}{\partial \xi} \bigg(\frac{1}{H}\frac{\partial \omega^n}{\partial \xi}\bigg), \\[3mm]\displaystyle
  \frac{\partial }{\partial t}\int_0^1 uH\,d\xi + \frac{\partial}{\partial x} \bigg(\int_0^1 u^2 H \, d\xi + \frac{\alpha h^2}{2}\bigg) =h- K\omega^n\big|_{\xi=0}, \quad \omega\big|_{\xi=1}=0,
 \end{array}
\end{equation}
where
\[ h=\int_0^1 H\,d\xi, \quad u=\int_0^\xi \omega H\,d\xi'. \]
The first equation in (\ref{eq:ConsForm}) is the local mass conservation law, the second and last equations correspond to the local and total horizontal momentum balance laws. 

Let us divide the interval $[0,1]$ into $M$ subintervals $0=\xi_0<\xi_1<...<\xi_M=1$ and introduce new variables
\begin{equation} \label{eq:zi-ui} 
 \begin{array}{l}\displaystyle
  z_i=\Phi(t,x,\xi_i), \quad u_i=u(t,x,\xi_i), \quad h_i=z_i-z_{i-1}, \\[3mm]\displaystyle
  \omega_i=\frac{u_i-u_{i-1}}{h_i}, \quad \bar{u}_i=\frac{u_i+u_{i-1}}{2}, \quad   Q=\sum_{i=1}^M \bar{u}_i h_i.
 \end{array} 
\end{equation}
Then we integrate Eq.~(\ref{eq:ConsForm}) with respect to $\xi$ over the intervals $(\xi_{i-1},\xi_i)$, $i=1,...,M$. Doing so, we take into account the equality $Hd\xi=dz$, boundary conditions $u_0=0$, $\omega_M=0$ and assume a piecewise linear approximation for horizontal velocity
\[ u(t,x,z)\approx \omega_i(t,x)(z-z_{i-1})+u_{i-1}, \quad z \in [z_{i-1},z_i]. \]
As a result, for $2M-1$ unknown functions $(h_1,...,h_M, \omega_2,...,\omega_{M-1}, Q)$, which are layers’ depths $h_i$, vorticities in the layers $\omega_i$, and a total fluid rate $Q$, we obtain the following system of balance laws:
\begin{equation} \label{eq:CL-diff}
 \begin{array}{l}\displaystyle
  \frac{\partial h_i}{\partial t}+ \frac{\partial}{\partial x}\big(\bar{u}_ih_i\big)=0, \quad (i=1,...,M) \\[3mm]\displaystyle
  \frac{\partial }{\partial t}\big(\omega_i h_i\big)+ \frac{\partial}{\partial x} \big(\bar{u}_i\omega_i h_i\big)=K \bigg(\frac{\omega_{i+1}^n-\omega_i^n}{h_i} -\frac{\omega_i^n-\omega_{i-1}^n}{h_{i-1}}\bigg), \quad (i=2,...,M-1) \\[3mm]\displaystyle
  \frac{\partial Q}{\partial t}+ \frac{\partial}{\partial x} \bigg(\sum\limits_{i=1}^M \Big(\bar{u}_i^2 h_i + \frac{\omega_i^2 h_i^3}{12}\Big) +\frac{\alpha h^2}{2}\bigg)= h-K\omega_1^n,
 \end{array}
\end{equation}
where
\[ \begin{array}{l}\displaystyle
    h=\sum\limits_{i=1}^M h_i,\quad \bar{u}_i=-\frac{\omega_i h_i}{2}+\sum\limits_{j=1}^i \omega_j h_j, \\[3mm]\displaystyle
    \omega_1=\frac{2}{(2h-h_1)h_1}\bigg(Q+\sum\limits_{i=2}^M h_i \Big(\frac{\omega_i h_i}{2}- \sum\limits_{j=2}^i \omega_j h_j\Big)\bigg). 
   \end{array} \]
To solve the differential conservation laws (\ref{eq:CL-diff}) numerically, one can apply standard methods based on various modifications of Godunov’s scheme. In our case, due to a large number of equations in system (\ref{eq:CL-diff}), it is convenient to use central schemes, which do not require exact or approximate solution of the Riemann problem. In this work we implement the Nessyahu--Tadmor second-order central scheme \cite{Nessyahu}. 

\section{Numerical simulation of roll waves}

In this Section we perform numerical modelling of the finite amplitude waves development from infinitesimal perturbation using both multilayer equations~(\ref{eq:CL-diff}) and depth-averaged model~(\ref{eq:av-model}). Let at the initial time $t=0$ the fluid depth $h=1$ and depth-averaged velocity $\bar{u}=1$. Note that this is a solution to Eq.~(\ref{eq:av-model}). For multilayer model~(\ref{eq:CL-diff}) at $t=0$ we choose $Q=1$, $z_i=i/M$, $u_i=f(z_i)$ ($i=0,...,M$) and according to formulae~(\ref{eq:zi-ui}) one can determine values of $h_i$ and $\omega_i$. We recall that $f(z)$ is the Nusselt-type profile given by (\ref{eq:Nus-profile}). We take the number of layers $M=20$. On the left boundary $x=0$ the constant flow rate is disturbed as follows
\begin{equation} \label{eq:perturb} 
  Q|_{x=0}=\bar{u}h|_{x=0}= 1+a_p\sin(\Omega t) 
\end{equation}
with $a_p=0.005$ and $\Omega=2\pi$. For unknown variables $\bU$ we set the following conditions $\bU_N=\bU_{N-1}$ on the right boundary of the computational domain $x=L$. Here $\bU_j$ is the value of vector-function $\bU$ in the nodal point $x_j$. An uniform grid with respect to the spatial variable $x$ is used for calculations, the number of nodes $N=8000$. The time step is determined by the Courant condition. 
\begin{figure}[t]
 \begin{center}
  \resizebox{.96\textwidth}{!}{\includegraphics{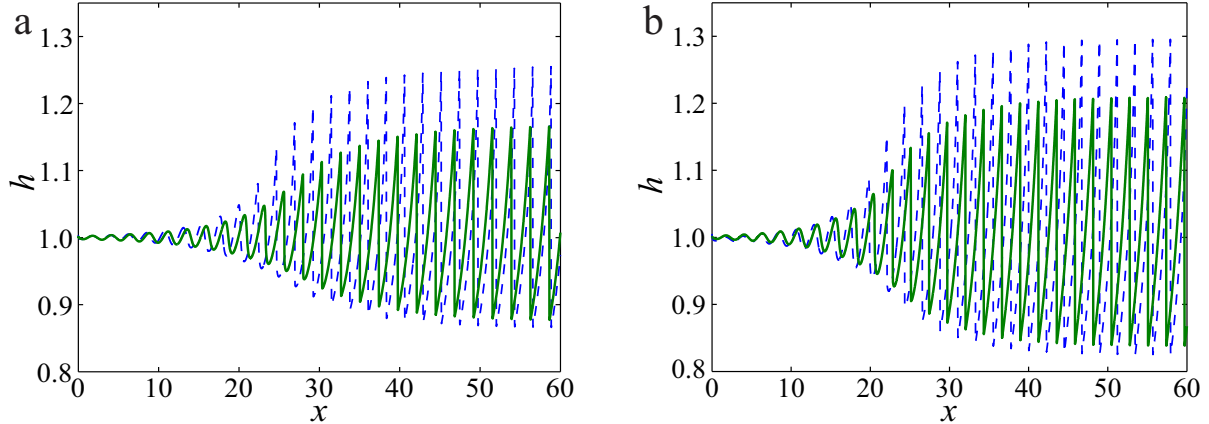}}\\[0pt]
  {\caption{Free surface $z=h$ at $t=35$, obtained by Eq.~(\ref{eq:av-model}) (solid curves) and Eq.~(\ref{eq:CL-diff}) (dashed curves) for $\alpha=1$. Power-law index $n=1$ (a) and $n=0.7$ (b).} \label{fig:fig_3}} 
 \end{center}
\end{figure}

\begin{figure}[t]
	\begin{center}
		\resizebox{.48\textwidth}{!}{\includegraphics{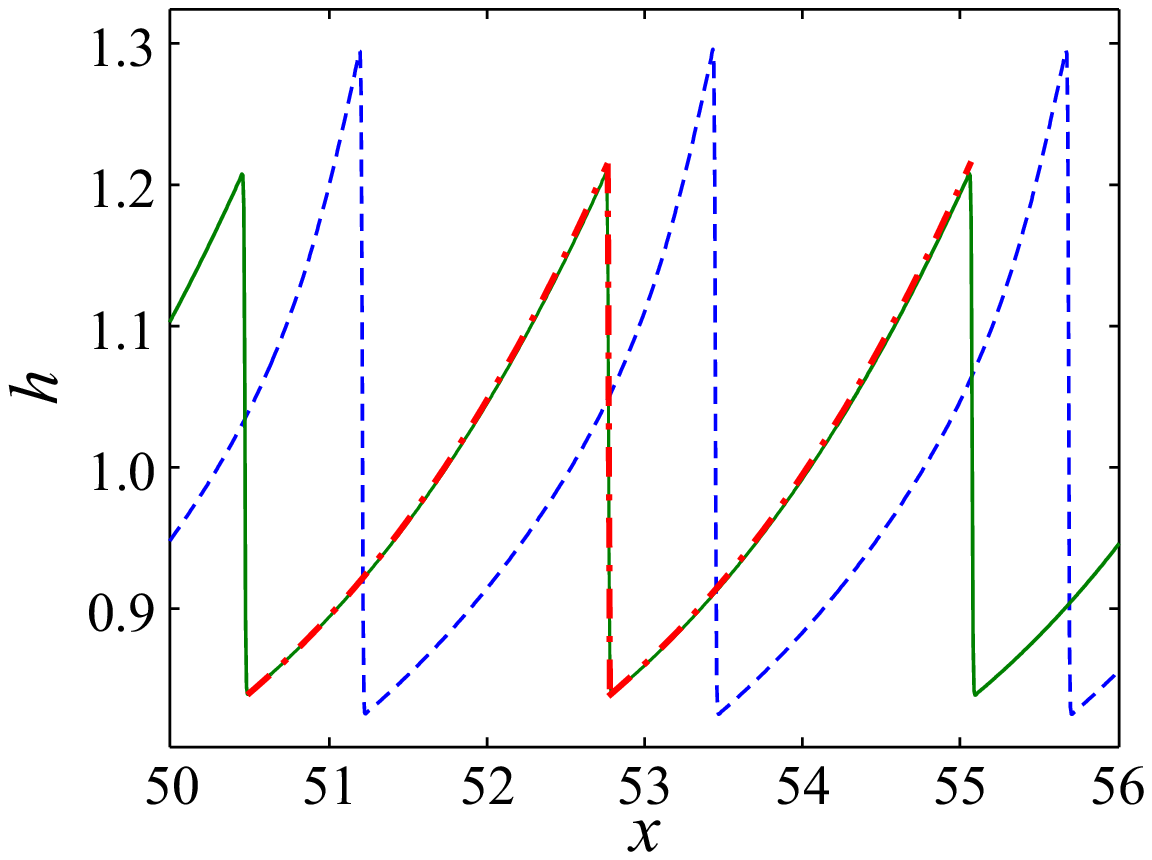}}\hfill
		\resizebox{.48\textwidth}{!}{\includegraphics{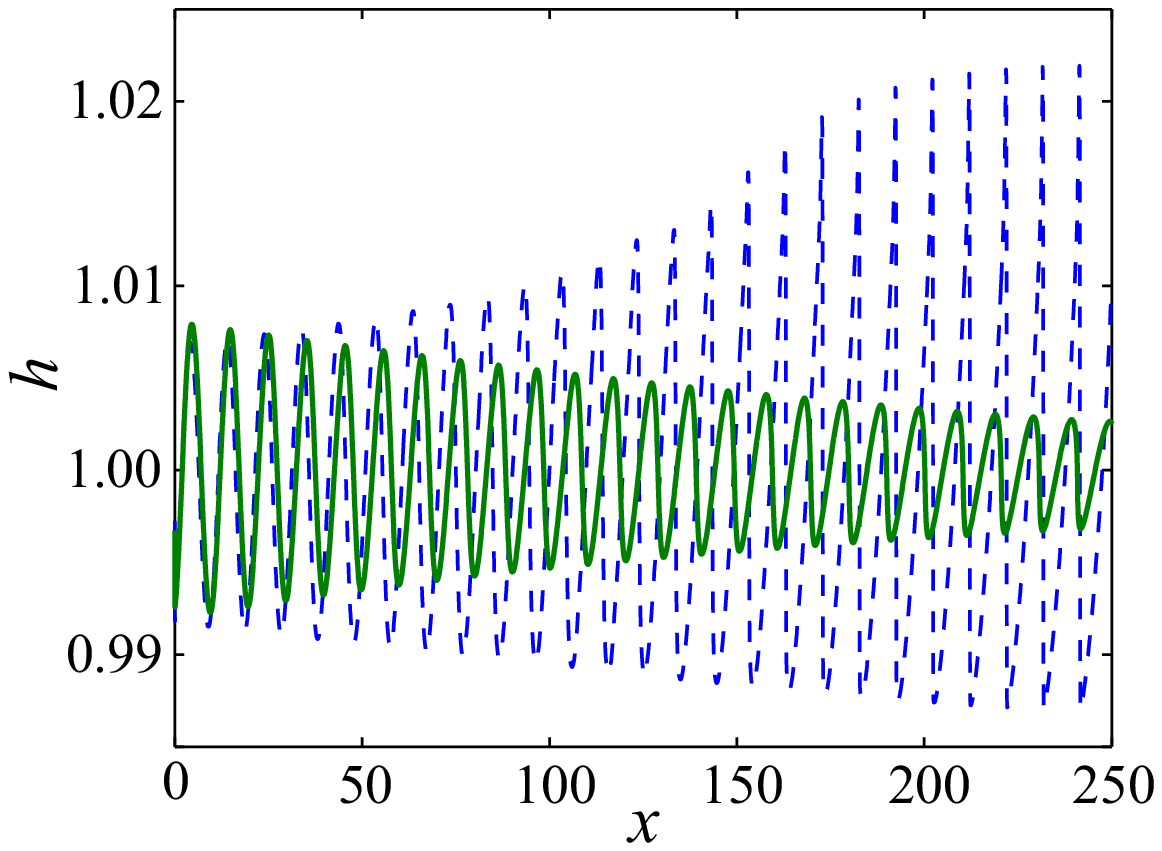}}\\[0pt]
		\parbox{.48\textwidth}{\caption{Part of Fig.~\ref{fig:fig_3}\,(b) on the interval $x\in (50,56)$; the dash-dotted curve corresponds to the exact solution of Eq.~(\ref{eq:av-model}). } \label{fig:fig_4}} \hfill
		\parbox{.48\textwidth}{\caption{Free surface $z=h$ at $t=90$ obtained by Eq.~(\ref{eq:av-model}) (solid curve) and Eq.~(\ref{eq:CL-diff}) (dashed curve) for $n=1$ and $\alpha=3.2$.} \label{fig:fig_5}} 
	\end{center}
\end{figure}

Let us take $\alpha=1$ and carry out calculation for different values of the power-law index $n\in [0.5,1]$. With this choice, the initial data correspond to the supercritical flow and the inequalities $\alpha<\alpha_n<\alpha_n^*$ hold, since $\alpha_n\geq 3$ and $\alpha_n^*\geq 3.5009$. Therefore, roll waves may occur. The results of calculations are shown in Fig.~\ref{fig:fig_3}. It can be observed that waves of finite amplitude are generated by small perturbations for the considered flow parameters. This is true both for model~(\ref{eq:av-model}) and for more general equations~(\ref{eq:thin-layer}) approximated by system~(\ref{eq:CL-diff}). Note that with a decrease in the fluid index $n$, perturbations develop more intensively. Fig.~\ref{fig:fig_4} corresponds to part of Fig.~\ref{fig:fig_3}\,(b) and demonstrates the form of the developed roll waves that consist of continuous flow regions connected by strong discontinuities. In this case the waves move with some positive velocity. In the continuity domain of the solution, a transition from a subcritical to a supercritical flow occurs in the frame moving with wave velocity. The dash-dotted curve in Fig.~\ref{fig:fig_4} corresponds to the exact solution of Eq.~(\ref{eq:av-model}) in the class of travelling waves that will be constructed in the next section. Note that the exact and numerical solutions almost completely coincide for the developed roll waves. 

\begin{table}[t]
\begin{center}
\caption{The relative differences between the amplitudes and wavelengths of roll waves obtained by Eq.~(\ref{eq:CL-diff}) and Eq.~(\ref{eq:av-model}).}\label{tab:tab_1}
\begin{tabular}{|c|c|c|c|c|c|c|} \hline
                   	& $n=1$   & $n=0.9$ & $n=0.8$ & $n=0.7$ & $n=0.6$ & $n=0.5$ \\ \hline
$\delta a\times 10$   & $2.644$ & $2.476$ & $2.295$ & $2.103$ & $1.997$ & $1.853$ \\ \hline 
$\delta l\times 10^2$ & $3.07$  & $3.03$  & $2.85$  & $2.52$  & $2.24$  & $2.05$ \\ \hline
\end{tabular}
\end{center}
\end{table}

The wave amplitude obtained by Eq.~(\ref{eq:CL-diff}) turns out to be higher than that calculated by model~(\ref{eq:av-model}), while the wavelengths practically coincide. The relative differences in the amplitudes $\delta a(n)=|a^*-a|/a^*$ and wavelengths $\delta l(n)$ for different values of $n$ are shown in Table~\ref{tab:tab_1}. Here $(a^*,l^*)$ and $(a,l)$ are the amplitude and wavelength of the roll waves according to Eq.~(\ref{eq:CL-diff}) and (\ref{eq:av-model}), respectively. In both cases, small perturbations are given in the form~(\ref{eq:perturb}) with $a_p=0.005$ and $\Omega=2\pi$. As follows from Table~\ref{tab:tab_1}, the differences between these waves become smaller with decreasing of the power-law index $n$. 

The probable reason for this is that for small values of $n$, the velocity profile weakly depends on the variable $z$ everywhere, with the exception of a small neighbourhood of the bottom (see Fig.~\ref{fig:fig_1}). That is why the calculation results using depth-averaged equation~(\ref{eq:av-model}) and multilayer approximation~(\ref{eq:CL-diff}) become closer for small $n$. We also note that the indicated features of the velocity profile complicate the calculations by Eq.~(\ref{eq:CL-diff}) for highly non-Newtonian fluid. This is due to the fact that the values of $\omega_i$ are close to zero in a significant number of layers and to perform calculations it is necessary to apply a more accurate resolution in the variable~$x$. 

As noted above, there is a difference between conditions~(\ref{eq:alpha-stab-gen}) and (\ref{eq:alpha-stab}), which ensure the stability of a Nusselt-type solution within the framework of models~(\ref{eq:av-model}) and (\ref{eq:thin-layer}), respectively. In particular, $\alpha_n=3$ and $\alpha_n^*\approx3.5$ at $n=1$. Let us take $\alpha_n<\alpha<\alpha_n^*$ and carry out numerical simulation of the evolution of small unsteady perturbations given on the left boundary based on Eq.~(\ref{eq:av-model}) and multilayer system~(\ref{eq:CL-diff}). We choose $\alpha=3.2$ and $n=1$. The calculation results at $t=90$ for perturbations (\ref{eq:perturb}) with parameters $a_p=0.1$ and $\Omega=0.6\pi$ are shown in Fig.~\ref{fig:fig_5}. As we can see in the figure, small perturbations of the basic solution are reduced when we use Eq.~(\ref{eq:av-model}). On the contrary, the same perturbations grow if we apply Eq.~(\ref{eq:CL-diff}). Thus, if the parameter $\alpha$ belongs to the interval $(\alpha_n,\alpha_n^*)$, then the results of calculations using two-dimensional equation~(\ref{eq:thin-layer}) and depth-averaged model~(\ref{eq:av-model}) are qualitatively different.

\section {Travelling waves}

The construction of a periodic piecewise-smooth solution of Eq.~(\ref{eq:av-model}), called roll waves, in the class of travelling waves was carried out in \cite{Ng_Mei}. Here we focus on the fact that this family of solutions is two-parameter and construct diagrams of roll waves existence, as well as compare the exact and numerical solutions. It should be noted that obtaining similar piecewise-smooth solutions for more general models~(\ref{eq:thin-layer}) or (\ref{eq:CL-diff}) is a rather complicated problem which requires a separate consideration.

The solutions to system~(\ref{eq:av-model}) in the class of travelling waves are determined from the equations
\begin{equation} \label{eq:tr_waves}
 \big((\bar{u}-D)h\big)'=0, \quad 
 \bigg((\beta\bar{u}-D)\bar{u}h +\frac{\alpha h^2}{2}\bigg)' = h- \bigg(\frac{\bar{u}}{h}\bigg)^n,
\end{equation}
where the prime denotes differentiation with respect to the variable $\zeta$, and $D$ is the
constant velocity of the travelling wave. Consider the case of $0<\bar{u}<D$. Then $(\bar{u}-D)h=-m$, where $m$ is a positive constant which is the discharge rate as seen by an observer moving at the wave speed $D$. Let us introduce the notation
\begin{equation} \label{eq:F-G-Delta}
 \begin{array}{l}\displaystyle
  G(h)= \big((\beta-1)Dh-\beta m\big) \bigg(D-\frac{m}{h}\bigg)+\frac{\alpha h^2}{2}\,, \\[3mm]\displaystyle
  \Delta(h)=\frac{dG}{dh}=(\beta-1)D^2- \frac{\beta m^2}{h^2}+ \alpha h, \quad 
  F(h)=h-\frac{1}{h^n}\bigg(D-\frac{m}{h}\bigg)^n.
 \end{array} 
\end{equation}
Then system~(\ref{eq:tr_waves}) reduces to the ordinary differential equation
\begin{equation}\label{eq:ODE}
 \frac{dh}{d\zeta}=\frac{F(h)}{\Delta(h)}\,.
\end{equation}

A characteristic feature of roll waves is the presence of a critical depth $y$ such that $\Delta(y)=0$. For $h=y$, a smooth transition occurs from a supercritical ($\Delta(h)<0$) to a subcritical ($\Delta(h)>0$) flow in the system of coordinates moving with the wave velocity $D$. To guarantee a smooth transition, one needs the condition $F(y)=0$ to be satisfied. In view of (\ref{eq:F-G-Delta}) we obtain the following representations for the parameters $m$ and $D$ in terms of the critical depth $y$:
\begin{equation}\label{eq:m-D} 
  m=Dy-y^{2+1/n}, \quad D=\beta y^{1+1/n}+\sqrt{\beta(\beta-1)y^{2(1+1/n)}+\alpha y}\,. 
\end{equation}
Note that the wave speed $D$ coincides with the `frozen' characteristic velocity $D=\lambda^+(\bar{u}(h),h)$ for $h=y$. 

\begin{figure}[t]
\begin{center}
\resizebox{.96\textwidth}{!}{\includegraphics{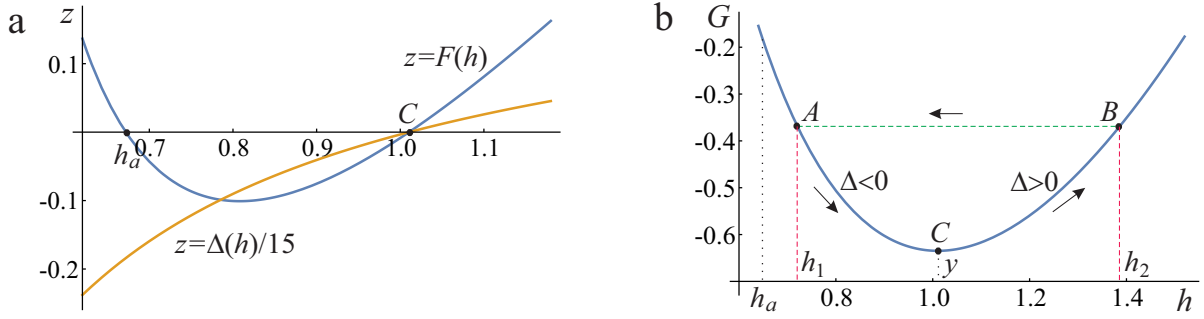}}\\[0pt]
{\caption{Mutual arrangement of the curves $z=F(h)$ and $z=\Delta(h)/15$ (a); graph of the function $G(h)$ (b). The graphs are obtained for $n=0.7$, $\alpha=1$ and $y=1.01$.} \label{fig:fig_6}} 
\end{center}
\end{figure}

Mutual arrangement of the curves $F(h)$ and $\Delta(h)$ from the right hand side of Eq.~(\ref{eq:ODE}) is shown in Fig.~\ref{fig:fig_6}\,(a). Such a qualitative arrangement of the curves takes place when the inequality 
\begin{equation}\label{eq:alpha-y}  
  \alpha<\frac{1+2n}{n^2}y^{1+2/n}
\end{equation}
is fulfilled. Indeed, $\Delta(h)$ is a monotonically increasing function ($\Delta'(h)>0$ for $h>0$) and $F(h)$ has two real roots $h=y$ and $h=h_a$ for $h>m/D$. In the interval between these roots, the function $F(h)$ is negative. Obviously, $h_a<y$ (as shown in the figure) if $F'(y)>0$ and $h_a>y$ if $F'(y)<0$. The derivative of the function $F(h)$ at the point $h=y$ is 
\[ F'(y)=1+n\big(2-Dy^{-1-1/n}\big). \]
In view of the second formula in (\ref{eq:m-D}), the inequality $F'(y)>0$ takes the form~(\ref{eq:alpha-y}). Note that this inequality coincides with Whitham's condition~(\ref{eq:Whitham}) for the existence of roll waves.

For given parameters $\alpha$ and $n$ of the model, one can construct a two-parameter family of periodic piecewise-smooth solutions as follows. First, we arbitrarily choose the critical depth $y>0$ so that the inequality~(\ref{eq:alpha-y}) holds. Then we choose the minimum fluid depth $h_1$ from the interval $(h_a,y)$. These are points $C$ and $A$ in Fig.~\ref{fig:fig_6}\,(b). Instead of $h_1$ one can choose the maximum fluid depth $h_2$ or the wave amplitude $a=h_2-h_1$. At $h=y$ the ratio $F/\Delta$ has the finite value because both the functions $F$ and $\Delta$ have the first order zeros at this point. For $h>h_1$ the right-hand side of Eq.~(\ref{eq:ODE}) is positive, so the function $h(\zeta)$ increases until the maximum value of $h_2$ is reached (point $B$ in the graph). The Hugoniot conditions at the shock front 
\begin{equation}\label{eq:Hugoniot}  
  (D-\bar{u}_1)h_1=(D-\bar{u}_2)h_2=m, \quad G(h_1)=G(h_2). 
\end{equation}
follow from the balance laws~(\ref{eq:av-model}). We denote the values of the functions on the right and left sides of the discontinuity by adding subscripts 1 and 2, respectively. The second relation in (\ref{eq:Hugoniot}) allows one to determine the maximum depth $h_2>h_1$. Thus, the periodic solution consists of a continuous interval $ACB$ and a shock $BA$ as shown in Fig.~\ref{fig:fig_6}\,(b). In this case the shock stability condition
\begin{equation}\label{eq:shock-cond} 
  \lambda^+(\bar{u}_1,h_1)<D<\lambda^+(\bar{u}_2,h_2) 
\end{equation}
is satisfied. 

A periodic piecewise-smooth solution of Eq.~(\ref{eq:av-model}) constructed according to the above algorithm for $n=0.7$, $\alpha=1$, $y=1.01$ and $h_1=0.84$ is shown Fig.~\ref{fig:fig_4} (dash-dotted curve). Using formulae~(\ref{eq:m-D}) and (\ref{eq:Hugoniot}), we find the maximum depth $h_2\approx 1.21$, the wave velocity $D\approx 2.30$ and $m\approx 1.29$. It can be seen from Fig.~\ref{fig:fig_4} that the constructed exact solution almost completely coincides with the non-stationary calculation performed according to Eq.~(\ref{eq:av-model}) for sinusoidal perturbations of a constant solution with frequency of $\Omega=2\pi$.

The construction of roll waves is impossible if $F'(y)<0$. As mentioned above, in this case $F(h)>0$, $\Delta(h)<0$ for $h\in (m/D,y)$ and $F(h)<0$, $\Delta(h)>0$ for $h\in (y, h_a)$. Therefore, the right-hand side in (\ref{eq:ODE}) is negative, i.e. the function $h(\zeta)$ is decreasing. Let us take $h_1$ from the interval $(y, h_a)$. Then we find $h_2<h_1$ from the Hugoniot condition~(\ref{eq:Hugoniot}). Eq.~(\ref{eq:ODE}) allows one to construct a continuous solution $h=h(\zeta)$ connecting the values of $h_1$ and $h_2$. However, it is impossible to pass from $h=h_2$ to $h=h_1$ using a shock wave, since the  condition~(\ref{eq:shock-cond}) is violated. Thus, in the case $F'(y)<0$, there is no periodic piecewise-smooth solution (roll waves).

\begin{figure}[t]
\begin{center}
\resizebox{.96\textwidth}{!}{\includegraphics{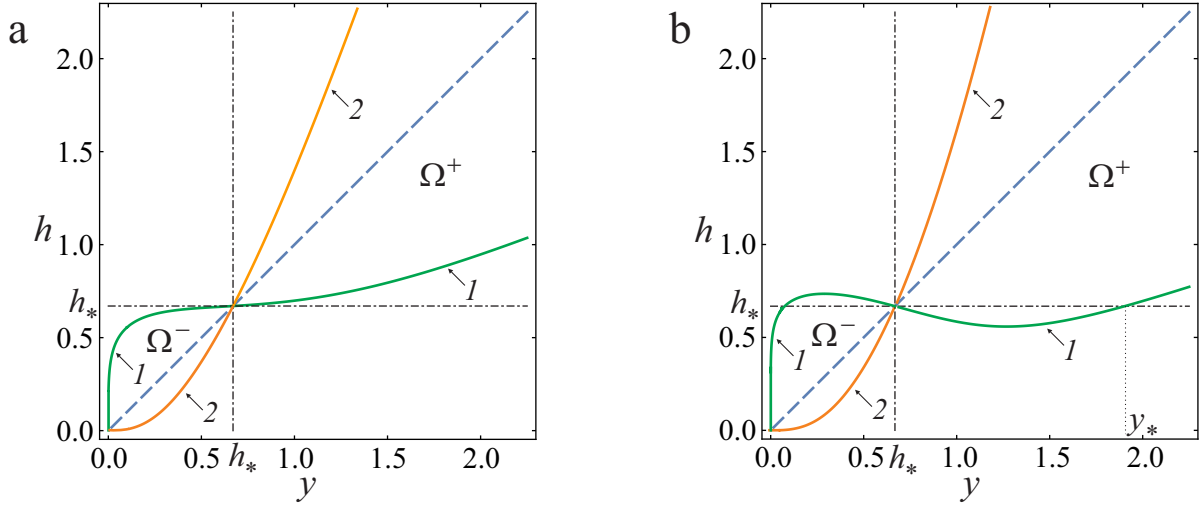}}\\[0pt]
{\caption{Roll wave diagrams for the parameters $\alpha=1$, $n=0.8$ (a) and $n=0.4$ (b). Curve 1 is given by equation $F(h,y)=0$ ($h\neq y$), curve 2 is conjugate. Dashed line is the diagonal $h=y$.} \label{fig:fig_7}} 
\end{center}
\end{figure}

The solution constructed above contains two parameters: the critical depth and the wave amplitude. Here we present a diagram of roll waves, which defines all possible points on the $(y,h)$-plane for which there exist roll waves. Let us take $\alpha=1$ (for other values $\alpha$ the qualitative behaviour of the diagrams does not change) and consider two values of $n$, namely $n=0.8>n_*$ and $n=0.4<n_*$, where $n_*=1/\sqrt{2}$. This choice is explained by the fact that \cite{Ng_Mei} noted some differences in the construction of roll waves for a power-law fluid with $n>n_*$ and $n<n_*$. To construct the diagram, we use the functions $G$, $F$, and $\Delta$ defined by formulae (\ref{eq:F-G-Delta}) and (\ref{eq:m-D}). These functions depend on variable $h$ and parameter $y$ in such a way that $\Delta(y,y)=0$ and $F(y,y)=0$. We find numerically the dependence $h=s(y)$ ($s\neq y$) at which the equation $F(s(y),y)=0$ is valid. This dependence is shown by curve 1 in Fig.~\ref{fig:fig_7}. Curve 2 in the same figure is conjugate to curve 1 and is obtained from the condition $G(s(y),y)=G(h,y)$ for $h\neq s(y)$. The equation $\Delta(h,y)=0$ has only one branch of solutions $h=y$ and does not impose additional restrictions on the range of admissible parameters. In the region bounded by curves 1 and 2, the ratio $F/\Delta$ does not change sign. However, for $y<h_*$ in the region $\Omega^-$ this ratio is negative and, as shown above, the construction of a piecewise-smooth periodic solution is impossible. The value $h_*$, defines the point of intersection of the `equilibrium' and `frozen' characteristic velocities, is given by formula~(\ref{eq:h_ast}). In the domain $\Omega^+$, the ratio $F/\Delta$ is positive. Thus, $\Omega^+$ defines all admissible values of the parameters of roll waves on the $(y,h)$-plane. 

For a fluid of flow index above $n_*$, the function $s(y)$ (such that $F(s(y),y)=0$) increases monotonically and $s(y)>h_*$ in the domain $\Omega^+$. For $n<n_*$, i.e. highly non-Newtonian fluid, the function $s(y)$ is non-monotonic and $s(y)<h_*$ on some interval $y\in(h_*,y_*)$. This means that it is possible to construct roll waves for which the minimum flow depth $h=h_1$ is less than the stability threshold $h=h_*$. However, the critical depth $h=y$ is higher than the specified threshold. 

\section{Conclusion}

In the paper we perform a nonlinear stability analysis on the film flow of a power-law fluid down an inclined plane by applying Whitham's criterion to the governing equations. To study the flow evolution, we use both 2D equations of the boundary layer theory~(\ref{eq:thin-layer}) and 1D depth-averaged model~(\ref{eq:av-model}). In dimensionless variables, these models contain two constants $n$ and $\alpha$, which are the power-law index and the flow parameter. It turn out that the velocities of the characteristics on a steady solution with a velocity profile of the Nusselt-type for model~(\ref{eq:thin-layer}) and the corresponding constant solution of equations~(\ref{eq:av-model}) are somewhat different (see Fig.~\ref{fig:fig_2}). Therefore, the stability criteria~(\ref{eq:alpha-stab}) and (\ref{eq:alpha-stab-gen}) for a steady flow of constant depth for the models under consideration do not completely coincide. This circumstance makes it possible to construct examples of flows for which small perturbations grow within the framework of Eq.~(\ref{eq:thin-layer}) and decay when we use depth-averaged Eq.~(\ref{eq:av-model}). However, this range of parameters in the $(n,\alpha)$-plane is rather narrow and, as a rule, both models give similar results of calculating the fluid depth and flow rate. 

The given numerical calculations confirm and illustrate the indicated above theoretical results. For the numerical simulation of the development of finite amplitude waves from infinitesimal disturbances on the surface, we use standard methods developed for solving systems of hyperbolic balance laws. The application of this approach to Eq.~(\ref{eq:av-model}) presents no difficulties. To solve long-wave Eq.~(\ref{eq:thin-layer}) by the same method, we derive 1D multilayer equations~(\ref{eq:CL-diff}) that approximate the original 2D system. The calculation results of the development of roll waves are shown in Fig.~\ref{fig:fig_3} and \ref{fig:fig_4}. As we can see, the amplitude of roll waves obtained by Eq.~(\ref{eq:CL-diff}) is slightly higher than calculated by Eq.~ (\ref{eq:av-model}), while the wavelength practically coincides. The qualitatively different evolution of small perturbations for the 2D and 1D models is presented in Fig.~\ref{fig:fig_5}. We also consider the construction of exact piecewise-smooth solutions of equations (\ref{eq:av-model}) in the class of travelling waves. We focus on the fact that this class of solutions is two-parameter. In the plane of parameters, the critical depth -- fluid depth, we construct diagrams of the region of roll waves existence for the given values of $n$ and $\alpha$ (see Fig.~\ref{fig:fig_7}).

The study shows that qualitative differences in the evolution of small perturbations on the solutions of the 2D and 1D models are manifested only in a narrow range of $n$ and $\alpha$. Outside this range of parameters, the use of the depth-averaged equations is justified and expedient.

\section*{Acknowledgements} 
This work is supported by Russian Foundation for Basic Research (Project No. 19-01-00498). The author thanks S.L. Gavrilyuk, V.Yu. Liapidevskii and I.V. Stepanova for fruitful discussions.

\renewcommand\baselinestretch{1}

\end{document}